\documentclass[fleqn,twoside]{article}
\usepackage{espcrc2}

\usepackage{graphicx}
\usepackage[figuresright]{rotating}

\newcommand{\AmS}{{\protect\the\textfont2
  A\kern-.1667em\lower.5ex\hbox{M}\kern-.125emS}}

\title{
 New algorithm of the high-temperature expansion for the Ising model\\
 in three dimensions
}

\author{H. Arisue\address[OPCT]{Osaka Prefectural College of Technology, 
   Saiwai-cho 26-12, Neyagawa, Osaka 572-8572, Japan}%
   \thanks{email address:arisue@las.osaka-pct.ac.jp}
     and
   T. Fujiwara\address[KU]{Faculty of General Studies, Kitasato University, 
   Kitasato 1-15-1, Sagamihara, Kanagawa 228-8555, Japan}
   \thanks{email address:fujiwara@clas.kitasato-u.ac.jp}
        }
       
\begin{document}
\begin{abstract}
 New algorithm of the finite lattice method is presented to generate 
the high-temperature expansion series of the Ising model. 
It enables us to obtain much longer series in three dimensions 
when compared not only to the previous algorithm of the finite lattice method 
but also to the standard graphical method. 
It is applied to extend the high-temperature series
of the simple cubic Ising model 
from $\beta^{26}$ to $\beta^{46}$ for the free energy 
and from $\beta^{25}$ to $\beta^{32}$ for the magnetic susceptibility.
\vspace{1pc}
\end{abstract}

\maketitle

\section{INTRODUCTION}
The finite lattice method\cite{Enting1977,Arisue1984,Creutz1991} is a powerful 
tool to generate the exact high- and low-temperature series and other exact 
expansion series for the spin models in the infinite volume limit.
In the graphical method, 
one has to list up all the relevant diagrams and count the number they appear.
In the finite lattice method we can skip these jobs and 
reduce the main task to the calculation of the partition 
function for the finite size lattices,
which can be done efficiently using the site-by-site 
integration\cite{Enting1980,Bhanot1990} without the graphical technique.

It has been extremely effective primarily in two dimensions,
but unfortunately it has worked out the series only in some limited cases 
in three dimensions.
This is because in two dimensions the CPU time and the computer memory 
needed to obtain the series to order $N$ increase exponentially with $N$, 
while 
they grow up exponentially with $N^2$ in three dimensions.
Here we present a new algorithm of the finite lattice method in which
the CPU time and the computer memory increase exponentially with $N\log{N}$.
It enables us to generate much longer series of the high-temperature expansion 
for the Ising model in three dimensions when compared not only to the previous 
algorithm of the finite lattice method but also to the graphical method. 

Although our main target is in three dimensions, the new algorithm 
of the finite lattice method applies in arbitrary dimensions,
so we describe it in two dimensions for convenience in the next two sections.

\section{FINITE LATTICE METHOD}
 In the finite lattice method to generate the high-temperature series
for the free energy in two dimensions 
we calculate the partition function 
$Z(l_x \times l_y)$ for the finite size lattices with 
$2(l_x + l_y) \le N$ and define recursively\cite{Arisue1984}
\begin{eqnarray}
\phi(l_x \times l_y)
&=&\log{[Z(l_x \times l_y)]} \nonumber\\
&&-\sum_{\scriptstyle 
  l_x^{\prime}\le l_x,l_y^{\prime}\le l_y,
       \atop\scriptstyle 
  l_x^{\prime} + l_y^{\prime} \ne l_x + l_y}
\phi(l_x^{\prime} \times l_y^{\prime})\;.
\end{eqnarray}
Here we use the notation for the lattice size such that the 
$1\times 1$ lattice means the unit square
composed of $2\times 2$ sites.
The Boltzmann factor for each bond 
is expressed as
\begin{equation}
  \exp\left(\beta s_k s_{k^{\prime}}\right)
 = \cosh\left(\beta\right) \left(1 + t s_k s_{k^{\prime}} \right), 
    \label{eqn:Boltzmann}
\end{equation}
with $\beta=J/k_BT$ and $t=\tanh\left(\beta\right)$.
We define the bond configuration as the set of bonds to which
the factor $t s_k s_{k^{\prime}}$ in (\ref{eqn:Boltzmann}) 
is assigned while the factor $1$ is assigned to the other bonds.
Non-vanishing contribution to the partition function comes only 
from the bond configuration in which the bonds form one or more closed loops.
Each of the closed loops is a polymer 
in the standard cluster expansion\cite{Muenster1981}.
Then the Taylor expansion of $\phi(l_x \times l_y)$ with respect to $t$ 
includes the contribution from all the clusters of polymers 
in the standard cluster expansion
that can be embedded into the lattice of $l_x \times l_y$ 
but cannot be embedded into any of its rectangular sub-lattices 
$l_x^{\prime} \times l_y^{\prime}$ and it starts from the 
term $t^{n}$ with $n=2(l_x+l_y)$,
which comes from the cluster of a single polymer (one closed loop of bonds)
that have two intersections with any line perpendicular 
to the lattice bonds. 
The expansion series of the free energy density in the infinite
volume limit is given by
\begin{equation}f=\sum_{2(l_x + l_y) \le N} \phi(l_x \times l_y)
\end{equation}

\section{NEW ALGORITHM}
In the standard algorithm of the finite lattice method 
the full partition function for the finite size lattice 
is calculated with all the bond configurations taken into account.
In order to obtain the series to a given order, however, 
it is enough to consider only a restricted number of the bond configurations.
Let us consider the anisotropic model of the simple cubic Ising model
with $\beta_i=J_i/k_BT$ and $t_i=\tanh{(\beta_i)}$ ($i=x,y$).
To obtain the series for 
$\phi(l_x \times l_y)$ to order $N_y=2l_y+\Delta N_y$
in $t_y$ we introduce\cite{Arisue2002a} in the new algorithm
$\phi(l_x \times l_y,\Delta N_y)$ 
defined recursively by 
\begin{eqnarray}
\phi(l_x \times l_y,\Delta N_y)
&=&\log{[Z(l_x \times l_y,\Delta N_y)]} \nonumber\\
&&\!\!\!\!\!\!\!\!\!\!\!\!\!\!\!\!\!\!\!\!\!\!\!\!\!\!\!\!\!\!\!\!\!\!\!\!
-\sum_{\scriptstyle 
l_x^{\prime}\le l_x,l_y^{\prime}\le l_y,
       \atop\scriptstyle 
l_x^{\prime} + l_y^{\prime}\ne l_x + l_y}
\phi(l_x^{\prime} \times l_y^{\prime},\Delta N_y)\;. \label{eqn:Zd}
\end{eqnarray}
Here the partition function 
$Z(l_x \times l_y,\Delta N_y)$ is calculated
only with the bond configurations taken into account 
that have orders $n_y{_i}$ in $t_y$ for the $i$-th layer perpendicular to the
y-direction satisfying 
\begin{equation}\sum_{i=1}^{l_y} \max(n_y{_i},2) \le 2l_y +\Delta N_y\;.
\end{equation} 

\begin{figure}[tbh!]
\setlength{\unitlength}{0.6mm}
\begin{picture}(110,50)
\put(10,0){\begin{picture}(50,60)(0,0)
\put(17,55){\mbox{(a)}}

\put(42, 2){\mbox{2}}
\put(42,12){\mbox{2}}
\put(42,22){\mbox{4}}
\put(42,32){\mbox{2}}
\put(42,42){\mbox{2}}
\multiput(0, 0)(2,0){20}{\line(1,0){1}}
\multiput(0,10)(2,0){20}{\line(1,0){1}}
\multiput(0,20)(2,0){20}{\line(1,0){1}}
\multiput(0,30)(2,0){20}{\line(1,0){1}}
\multiput(0,40)(2,0){20}{\line(1,0){1}}
\multiput(0,50)(2,0){20}{\line(1,0){1}}
\multiput( 0,0)(0,2){25}{\line(0,1){1}}
\multiput(10,0)(0,2){25}{\line(0,1){1}}
\multiput(20,0)(0,2){25}{\line(0,1){1}}
\multiput(30,0)(0,2){25}{\line(0,1){1}}
\multiput(40,0)(0,2){25}{\line(0,1){1}}
\thicklines
\multiput(10, 0)(10,0){1}{\line(1,0){10}}
\multiput( 0,10)(10,0){1}{\line(1,0){10}}
\multiput(20,10)(10,0){2}{\line(1,0){10}}
\multiput(10,20)(20,0){2}{\line(1,0){10}}
\multiput(20,30)(20,0){1}{\line(1,0){10}}
\multiput( 0,50)(20,0){1}{\line(1,0){10}}

\multiput( 0,10)(0,10){4}{\line(0,1){10}}
\multiput(10,00)(0,10){1}{\line(0,1){10}}
\multiput(10,20)(0,10){3}{\line(0,1){10}}
\multiput(20,00)(0,10){1}{\line(0,1){10}}
\multiput(20,20)(0,10){1}{\line(0,1){10}}
\multiput(30,20)(0,10){1}{\line(0,1){10}}
\multiput(40,10)(0,10){1}{\line(0,1){10}}
\end{picture}
}
\put(70,0){\begin{picture}(50,50)(0,0)
\put(17,55){\mbox{(b)}}

\put(42, 2){\mbox{0}}
\put(42,12){\mbox{4}}
\put(42,22){\mbox{4}}
\put(42,32){\mbox{0}}
\put(42,42){\mbox{2}}
\multiput(0, 0)(2,0){20}{\line(1,0){1}}
\multiput(0,10)(2,0){20}{\line(1,0){1}}
\multiput(0,20)(2,0){20}{\line(1,0){1}}
\multiput(0,30)(2,0){20}{\line(1,0){1}}
\multiput(0,40)(2,0){20}{\line(1,0){1}}
\multiput(0,50)(2,0){20}{\line(1,0){1}}
\multiput( 0,0)(0,2){25}{\line(0,1){1}}
\multiput(10,0)(0,2){25}{\line(0,1){1}}
\multiput(20,0)(0,2){25}{\line(0,1){1}}
\multiput(30,0)(0,2){25}{\line(0,1){1}}
\multiput(40,0)(0,2){25}{\line(0,1){1}}
\thicklines
\multiput( 0,10)(10,0){2}{\line(1,0){10}}
\multiput(30,10)(10,0){1}{\line(1,0){10}}
\multiput(10,20)(10,0){3}{\line(1,0){10}}
\multiput(20,30)(20,0){1}{\line(1,0){10}}
\multiput( 0,30)(20,0){1}{\line(1,0){10}}

\multiput( 0,10)(0,10){2}{\line(0,1){10}}
\multiput(10,20)(0,10){1}{\line(0,1){10}}
\multiput(20,10)(0,10){2}{\line(0,1){10}}
\multiput(30,10)(0,10){2}{\line(0,1){10}}
\multiput(40,10)(0,10){1}{\line(0,1){10}}

\multiput( 0,40)(10,0){3}{\line(1,0){10}}
\multiput( 0,50)(10,0){3}{\line(1,0){10}}

\multiput( 0,40)(0,10){1}{\line(0,1){10}}
\multiput(30,40)(0,10){1}{\line(0,1){10}}
\end{picture}
}
\end{picture}
\vspace{-0.5cm}
\caption{\label{fig:config2d}
Examples of the bond configurations.}
\end{figure}
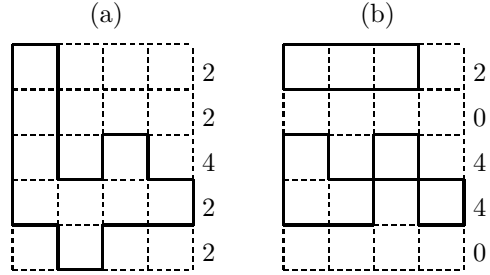

We neglect every bond configuration for the partition function
that has $\sum_{i=1}^{l_y} \max(n_y{_i},2) > 2l_y +\Delta N_y$
among the configurations that have 
$\sum_{i=1}^{l_y} n_y{_i} \le 2l_y +\Delta N_y$.
It is easy to prove that any of the neglected configuration
does not contribute to $\phi(l_x\times l_y)$ in the order 
lower than or equal to $N_y = 2l_y+\Delta N_y$.
For such a configuration at least one of the $n_y{_i}$'s should be zero, 
so they are disconnected configuration(composed of more than one polymer)
or they can be embedded into a rectangular sub-lattice of 
$l_x^{\prime} \times l_y^{\prime}$ with $l_y^{\prime} < l_y$
and in either case they do not contribute to $\phi(l_x \times l_y)$ 
in the order lower than or equal to $N_y = 2l_y +\Delta N_y$. 
They contribute to $\phi(l_x \times l_y)$ only 
in higher order than $N_y = 2l_y +\Delta N_y$
by constituting the connected cluster of polymers
together with the polymers coming from other configurations 
that have $n^{\prime}{_y{_i}}\ge 2$ for the layer $i$ with $n_y{_i}=0\;$.
Examples of the bond configurations are shown in Figure~\ref{fig:config2d}
for $l_x=4,\ l_y=5$.
The example (a) has $\{n_y{_i}\}=\{2,2,4,2,2\}$
and should be taken into account for $\Delta N_y=2$, while 
the example (b) has $\{n_y{_i}\}=\{0,4,4,0,2\}$
and should be neglected for the same $\Delta N_y=2$, 
in spite of the fact that the total order of (b) in $t_y$ is smaller than 
$2l_y+\Delta N_y$.

The contribution of the bond configuration with $\{n_y{_i}\}$
to the partition function of the finite size lattice 
can be calculated by the transfer matrix formalism as
\begin{equation}Z(\{n_y{_i}\})=V_{0,j_1}t_y^{n_y{_1}}V_{j_1,j_2}t_y^{n_y{_2}}
 \cdots t_y^{n_y{_{l_y}}} V_{j_{l_y},0}\,. \label{eqn:Zv}
\end{equation}
Here $V_{j_i,j_{i+1}}$ is the transfer matrix element 
with incoming $n_y{_i}$ spins and outgoing $n_y{_{\;i+1}}$ spins.
The summations over the spin locations $j_1,j_2,\cdots$ 
of the $n_y{_1}, n_y{_2}, \cdots$ spins, respectively, 
are assumed in the right hand side of (\ref{eqn:Zv}). 

In three dimensions, which is our main target, 
the transfer matrix element itself can be regarded as the partition function 
in two dimensions and can be calculated efficiently 
by the site-by-site construction\cite{Enting1980,Bhanot1990}.
In the new algorithm most of the CPU time should be used 
for the calculation of these transfer matrix elements,
and the total CPU time to generate the series to order $N$ can be estimated 
to increase exponentially with the leading term of the exponent proportional 
to $N\log N$.

\section{RESULT}
The new algorithm is applied to generate the high-temperature series 
of the simple cubic Ising model 
for the free energy\cite{Arisue2002a} to $\beta^{46}$ 
and for the magnetic susceptibility\cite{Arisue2002c} to $\beta^{32}$. 
The obtained series agree with the previous series to $\beta^{26}$
for the  free energy\cite{Guttmann1994}, which was obtained 
by the previous algorithm of the finite lattice method,
and to $\beta^{25}$ for the susceptibility\cite{Butera2002},
which was obtained by the graphical method.
It should be commented that the previous algorithm of the finite lattice method
can generate the susceptibility series only to $\beta^{13}$.

Preliminary analysis of the free energy series 
using the inhomogeneous differential approximation and the ratio method 
biased by the value of the critical point $\beta_c$
gives the estimation of the 
critical exponent for the specific heat as 
$\alpha=0.104(1)$ and $\alpha=0.108(1)$ 
corresponding to the result of $\beta_c=0.22165459(10)$
and $\beta_c=0.2216595(15)$,respectively, 
of the recent two Monte Carlo simulations\cite{Blote1999,Ito2000}.
Unfortunately the free energy series can give the estimate 
of the critical point itself only in poor precision.
The analysis of the susceptibility series 
by the ratio method gives the estimate of the critical point 
as $\beta_c=0.2216550(5)$ and the critical exponent as $\gamma=1.2370(2)$.
This estimation for the critical exponent do not use 
the value of the critical point as the input. 
It is consistent with the recent estimate $\gamma=1.2371(4)$ 
obtained from the high-temperature series of the generalized Ising model
\cite{Campostrini2002}. 

\section{DISCUSSION}
We have presented the new algorithm of the finite lattice method 
for the high-temperature expansion of the Ising model.
It has been applied to the high-temperature expansion of the free energy
and the magnetic susceptibility for the simple cubic Ising model.
It can be applied to the high-temperature expansion of other quantities 
such as the correlation length
and it can also be applied to the models with continuous spin variables 
such as the XY model in three dimensions.
We note that the dimensionality of the lattice is not restricted to three 
as can seen by the fact that the description of the new algorithm in section 3
was given in two dimensions.
Furthermore the basic idea of the new algorithm 
can be used in the low-temperature expansion 
for the spin models.
We can expect that the new algorithm will enable us 
to generate the series for these models that are much longer
than the presently available series.

\end{document}